\documentclass[11pt,prd,nofootinbib,preprint]{revtex4}
\usepackage[a4paper, total={7in, 10.5in}]{geometry}
\usepackage{bm}
\usepackage{color}
\usepackage{xcolor}
\usepackage{float}
\usepackage{booktabs}
\usepackage{multirow}
\usepackage{amsmath}
\usepackage{graphicx}
\usepackage{esint}
\usepackage{epsfig}
\usepackage{caption}
\usepackage{subcaption}
\usepackage{hyperref}
\usepackage{cancel}
\usepackage{array}
\usepackage{multirow}
\usepackage{makecell}

\newcommand{\ua}{$\uparrow$}
\newcommand{\da}{$\downarrow$}
\usepackage{slashed}
\usepackage{appendix}
\usepackage{tabularx}
\usepackage{pdfpages}
\usepackage{epstopdf}
\usepackage{siunitx}
\usepackage{soul}
\usepackage{cancel}
\usepackage[normalem]{ulem}
\makeatletter

\DeclareFontEncoding{LGR}{}{}

\ProvideTextCommand{\~}{LGR}[1]{\char126#1}

\newcommand{\lyxmathsym}[1]{\ifmmode\begingroup\def\b@ld{bold}
  \text{\ifx\math@version\b@ld\bfseries\fi#1}\endgroup\else#1\fi}

\@ifundefined{textcolor}{}
{%
\definecolor{BLACK}{gray}{0}
 \definecolor{WHITE}{gray}{1}
 \definecolor{RED}{rgb}{1,0,0}
 \definecolor{GREEN}{rgb}{0,1,0}
 \definecolor{BLUE}{rgb}{0,0,1}
 \definecolor{CYAN}{cmyk}{1,0,0,0}
 \definecolor{MAGENTA}{cmyk}{0,1,0,0}
 \definecolor{YELLOW}{cmyk}{0,0,1,0}
}

\usepackage[T1]{fontenc}
\begin{document}
\begin{center}
{\bf\Large\boldmath
%Global fit of B anomalies with implications via  $\Lambda_{b}\rightarrow\Lambda_{c}\tau\bar{\nu}_{\tau}$ five-fold distributions
Weak decay of the positronium ion
}\\[5mm]
\par\end{center}
\begin{center}
\setlength{\baselineskip}{0.2in} {Nishat Ul Sani $^{a,}$\footnote{nishat.sani764@gmail.com},
Muhammad Jamil Aslam $^{a,}$\footnote{jamil@qau.edu.pk}, and
Ishtiaq Ahmed$^{b,}$\footnote{ishtiaqmusab@gmail.com}  
}\\[5mm] $^{a}$~\textit{Department of Physics, Quaid-i-Azam
University, Islamabad 45320, Pakistan.}\\
 $^{b}$~\textit{National Center for Physics, Quaid-i-Azam University Campus, Islamabad 44000, Pakistan.}\\[5mm]
% $^{c}$~\textit{Department of Physics, International Islamic University,
%Islamabad 44000, Pakistan.}\\[5mm] 
\par\end{center}
\begin{abstract}
The positronium ion ($\mathrm{Ps}^-$), a coulombic three-body bound state of two electrons and a positron, predominantly decays via electron-positron annihilation into electromagnetic final states. While its radiative decay channels have been extensively studied, much less attention has been given to weak processes in this system. In this work, we investigate the rare decay $\mathrm{Ps}^- \to e^- \nu_\mu \bar{\nu}_\mu$, obtained by replacing the photon in $\mathrm{Ps}^- \to e^- \gamma$ with a virtual $Z$ boson. Treating the three-body process as an effective two-body transition, $\mathrm{Ps}^- \to e^- Z^*\left(\to \nu_\mu\bar{\nu}_\mu\right)$, we compute the decay rate by explicitly evaluating all spin configurations of the initial bound state and final particles. The result agrees with that obtained using the standard spin-summation formalism of quantum field theory. We find that the branching ratio is comparable to that of the weak decay of ortho-positronium, $\mathrm{o\text{-}Ps} \to \gamma \nu \bar{\nu}$.
\end{abstract}

%\date{\today}
\maketitle

\section{Introduction}\label{sec1}
The possibility for any particle to decay can be determined by the kinematics of the decay and by discrete symmetries. These symmetries give rise to the selection rules that determine whether a particular decay is allowed~\cite{Peccei}. A well-known example of a two-particle leptonic bound state is Positronium (Ps)~\cite{Czarnecki:1999uk}, a hydrogen-like bound state of an electron and a positron having spin singlet (para-positronium) or triplet (ortho-positronium). For these states, all even- and odd-number photon decay modes are allowed in quantum electrodynamics (QED)~\cite{Sen2019, Ore1949}, except for a single-photon decay, which is kinematically forbidden. However, the single-photon decay becomes allowed via the weak interactions due to the violation of parity ($\mathcal{P}$) and charge conjugation ($\mathcal{C}$). One such process is $\mathrm{Ps}\to\gamma\nu\bar{\nu}$~\cite{Pokraka:2016jgy, love2015}, where the decay rate of ortho-positronium into a photon and a $\nu_\mu\bar{\nu}_\mu$ pair is approximately three orders of magnitude larger than the corresponding para-positronium decay rate \cite{Pokraka:2016jgy}.

It is worth noting that this symmetry restriction is lifted when an additional charged particle $\left(e^\pm\right)$ accompanies the $\mathrm{Ps}$ atom in the initial state, as in the positronium ion $\mathrm{Ps}^{\pm}$.
It was first observed in 1981~\cite{PhysRevLett.46.717}, and efficient production methods were subsequently developed~\cite{Nagashima_2008,NAGASHIMA201495}.
To form a three-leptons bound state, the two electrons are in a spin-singlet configuration $\left(
\frac{\uparrow\downarrow - \downarrow\uparrow}{\sqrt{2}}\right)$, ensuring that the total wave function of $\mathrm{Ps}^-$ is antisymmetric.

In addition,
The presence of the additional charged constituent allows the otherwise forbidden QED one-photon annihilation channel, $\mathrm{Ps}^{-} \to e^{-}\gamma$~\cite{PhysRev.44.510.2}, since the spectator particle can recoil after absorbing or emitting one of the photons and ensure overall energy–momentum conservation. 

A complete calculation of the electromagnetic decay, $\mathrm{Ps}^-\to e^-\gamma$, including all leading-order (LO) Feynman diagrams, was first carried out by Kryuchkov~\cite{Kryuchkov_1994}, refining earlier results reported in Refs.~\cite{Ho1983, Chu1986}. The result was later rederived in Ref.~\cite{Aslam:2021uqu} using a more direct method, in which the total amplitude is constructed from spinor bilinears along the two fermion lines, expressed in terms of Dirac $\gamma$-matrices and combined with the standard QED vertex and propagator factors.

Despite the extensive literature on the composite positronium systems within the framework of QED, their weak-interaction decay channels have received comparatively little attention. This omission is particularly relevant for the negative positronium ion, $\mathrm{Ps}^-$, whose electroweak decay offers a clean probe of purely three-body  leptonic weak interactions in a bound-state environment.  In this work, we investigate the rare electroweak decay $\mathrm{Ps}^- \to e^- \nu_\mu \bar{\nu}_\mu$.

Compared with the electroweak decay $\mathrm{Ps}\to \gamma\nu_\mu\bar{\nu}_\mu$, the decay $\mathrm{Ps}^-\to e^-\nu_\mu\bar{\nu}_\mu$ exhibits two distinctive features. First, whereas positronium annihilates from either the para-positronium $(S=0)$ or ortho-positronium $(S=1)$ state, both spin configurations contribute in $\mathrm{Ps}^-$, leading to interference effects in the decay rate. Second, the presence of a massive electron in the final state gives rise to a phase-space structure that differs significantly from that of $\mathrm{Ps}\to \gamma\nu_\mu\bar{\nu}_\mu$, where all final-state particles are massless. These features make $\mathrm{Ps}^-\to e^-\nu_\mu\bar{\nu}_\mu$ a nontrivial extension of the positronium case.
Motivated by these distinctive features, we investigate the decay $\mathrm{Ps}^- \to e^- \nu_\mu \bar{\nu}_\mu $ and evaluate its decay rate using two complementary theoretical approaches.
\\
In the first approach, we treat the process as a factorized transition,
$\mathrm{Ps}^- \to e^- Z^*$ followed by $Z^* \to \nu_\mu \bar{\nu}_\mu$, and express the full three-body decay rate in terms of an effective two-body rate as done for $\mathrm{Ps}\to \gamma\nu\bar{\nu}$ in \cite{Pokraka:2016jgy}. We compute the corresponding amplitudes for all possible initial spin configurations, including the eight LO Feynman diagrams contributing to the process. By integrating over the virtual $Z$ boson phase space, we recover the complete three-body decay rate for $\mathrm{Ps}^- \to e^- \nu_\mu \bar{\nu}_\mu$. 

In the second approach, we perform a direct calculation of the three-body decay using the standard spin-summation technique of Quantum Field Theory (QFT), evaluating the squared amplitude via trace methods (Casimir trick). We demonstrate that the two methods yield identical results. Similar consistency between two approaches has already been established for multi-lepton bound systems, such as dipositronium ($\mathrm{Ps}_2$)~\cite{Munir:2023gsa} and leptons-hadron bound state, \textit{i.e.}, the positronium hydride ($\mathrm{PsH}$)~\cite{PhysRevA.109.032820}. We further determine the branching ratio of the decay and compare it with the corresponding electroweak decay of positronium~\cite{Pokraka:2016jgy}. 

The work is organized as follows. In Sec.~\ref{weak-decay-Ps-ion}, we present the leading-order Feynman diagrams and compute the decay rates for specific spin configurations, as well as the results obtained via the spin-summation method. Section~\ref{conclusions} summarizes our findings. Details of the phase-space decomposition, decay kinematics, and spinor structures expressed in terms of $\gamma$-matrices are provided in the Appendix \ref{app:phase}.

\section{Weak Decay of Positronium ion to electron and neutrinos}\label{weak-decay-Ps-ion}
\begin{figure}[h!]
    \centering
    
    \begin{subfigure}[b]{0.22\textwidth}
        \centering
        \includegraphics[width=\textwidth]{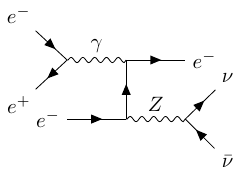}
        \caption{}
        \label{fig:f1}
    \end{subfigure}%
    \begin{subfigure}[b]{0.22\textwidth}
        \centering
        \includegraphics[width=\textwidth]{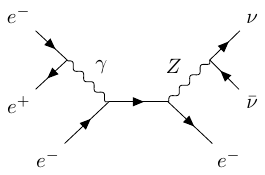}
        \caption{}
        \label{fig:f2}
    \end{subfigure}%
    \begin{subfigure}[b]{0.22\textwidth}
        \centering
        \includegraphics[width=\textwidth]{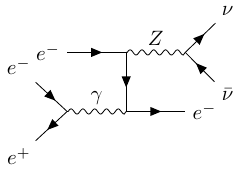}
        \caption{}
        \label{fig:f3}
    \end{subfigure}%
    \begin{subfigure}[b]{0.22\textwidth}
        \centering
        \includegraphics[width=\textwidth]{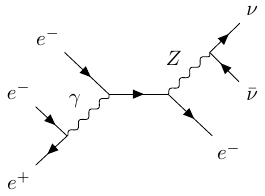}
        \caption{}
        \label{fig:f4}
    \end{subfigure}
    
    \vspace{0.3cm} % vertical space between rows

    \begin{subfigure}[b]{0.22\textwidth}
        \centering
        \includegraphics[width=\textwidth]{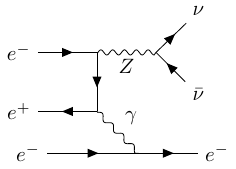}
        \caption{}
        \label{fig:f5}
    \end{subfigure}%
    \begin{subfigure}[b]{0.22\textwidth}
        \centering
        \includegraphics[width=\textwidth]{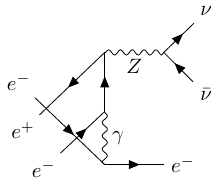}
        \caption{}
        \label{fig:f6}
    \end{subfigure}%
    \begin{subfigure}[b]{0.22\textwidth}
        \centering
        \includegraphics[width=\textwidth]{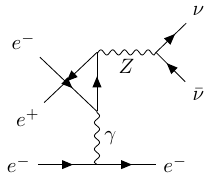}
        \caption{}
        \label{fig:f7}
    \end{subfigure}%
    \begin{subfigure}[b]{0.22\textwidth}
        \centering
        \includegraphics[width=\textwidth]{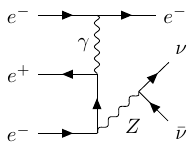}
        \caption{}
        \label{fig:f8}
    \end{subfigure}
    
    \caption{The leading- order (LO) Feynman diagrams for $\mathrm{Ps}^-\to e^-\nu_\mu\bar{\nu}_\mu$ weak decay.}
    \label{fig:dia}
\end{figure}

In Standard Model (SM), the relevant LO Feynman diagrams for the decay 
$\mathrm{Ps}^- \to e^- \nu_\mu \bar{\nu}_\mu$ are shown in Fig.~\ref{fig:dia}. 
This decay proceeds via the annihilation of an \(e^- e^+\) pair in the presence 
of a third (spectator) electron. The annihilation produces a virtual $Z$ boson, which subsequently decays into a neutrino-antineutrino pair, 
$\nu_\mu \bar{\nu}_\mu$. In addition, a photon emitted in the annihilation 
process is absorbed by the spectator electron.

One possibility is the single-photon annihilation of the $e^-e^+$ pair 
in the spin-triplet state. The emitted photon is absorbed by the remaining 
electron either before or after the emission of the virtual $Z$ boson, 
as illustrated in Fig.~\ref{fig:dia} (a-d). Alternatively, the annihilating 
\(e^- e^+\) pair may be in a spin-singlet state, as shown in 
Fig.~\ref{fig:dia} (e--h). This contribution is referred to as the radiationless
decay channel and corresponds to the coalescence of all three particles at a 
single spatial point.

To account for this configuration, we include the factor \(\Psi(0,0,0)\) in the 
scattering amplitude, which denotes the \(\mathrm{Ps}^-\) wave function evaluated at the origin. Its modulus squared, \(|\Psi(0,0,0)|^2\), is 
equal to the expectation value of the product of two Dirac delta functions 
representing the spatial coincidence of the three particles~\cite{Frolov2007}:
\begin{align}
	|\Psi(0,0,0)|^2=\langle \delta^3(r_{e^+e^-})\delta^3(r_{e^-e^-})\rangle=\langle \delta_{+--}\rangle a_B^{-6}\approx3.589 \times 10^{-5}\alpha^6m^6\;,\label{coel-prob}
\end{align} 
where $m$ is the mass of $e^-\left(e^+\right)$, $\alpha$ is the fine-structure constant, and $a_B=\frac{1}{\alpha m}$ is the Bohr radius.

We assume the initial incoming four-momenta of $e^-$, $e^+$, and $e^-$ as $p_1,p_2$, and $p_3$ while outgoing four-momenta are denoted by $k_i$ (${i=1,2,3}$), where $k_1$ is the four-momentum of neutrino, $k_2$ the antineutrino, and $k_3$ the $e^-$. Since the binding energy of Ps$^-$ is 0.326 eV~\cite{czarnecki2012}, which is small compared to the rest mass of initial state particles, we take initial leptons to be at rest with $p_1=p_2=p_3=(m,0,0,0)$. Furthermore, in the SM the neutrinos are considered to be massless.

Using off-shell factorization, the three-body decay rate for 
$\mathrm{Ps}^- \to e^- \nu_\mu \bar{\nu}_\mu$ 
can be expressed in terms of the two-body decay rate 
$\mathrm{Ps}^- \to e^- Z^*$, in close analogy to the treatment of 
$\mathrm{Ps} \to \gamma \nu \bar{\nu}$ in Ref.~\cite{Pokraka:2016jgy}. 
To this end, we first construct the three-body amplitude for 
$\mathrm{Ps}^- \to e^- \nu_\mu \bar{\nu}_\mu$ 
and factorize the corresponding three-body phase space into a product of two two-body phase spaces, 
$\mathrm{Ps}^- \to e^- Z^*$ and $Z^* \to \nu_\mu \bar{\nu}_\mu$, 
as detailed in Appendix~\ref{app:phase}. 
After integrating over the $\nu_\mu \bar{\nu}_\mu$ phase space, 
the decay rate for 
$\mathrm{Ps}^- \to e^- \nu_\mu \bar{\nu}_\mu$ 
can be expressed in terms of the decay rate for 
$\mathrm{Ps}^- \to e^- Z^*$ (see Appendix~\ref{app:3bodydecay}):
\begin{align}
	\Gamma_{\mathrm{Ps}^- \to e^- \nu_\mu\bar{\nu}_\mu}=\frac{4G_F^2}{2\pi^2\alpha}\int \frac{dq^2}{2\pi}q^2\Gamma_{\mathrm{Ps}^- \to e^- Z^*}\;,
\end{align}
where $G_F \simeq 1.166 \times 10^{-5}\,\mathrm{GeV}^{-2}$ denotes the Fermi constant~\cite{Marciano1999}, and $q=k_1+k_2$ is the  four-momentum of the virtual $Z-$boson, or the total momentum of neutrino pair. 

For the two-body process $\mathrm{Ps}^- \to e^- Z^*$, we work in the $\mathrm{Ps}^-$ rest frame and choose the coordinate system such that the outgoing electron propagates along the $+z$ axis, \textit{i.e.},  its four-momentum is then given by $
k_3 = \left(E_e,\, 0,\, 0,\, \left|\vec{k}_3\right|\right)$.
Momentum conservation requires the virtual gauge boson $Z^*$ to recoil along the $-z$ axis with four-momentum $
q = \left(E_q,\, 0,\, 0,\, -\left|\vec{k}_3\right|\right)$.
With this kinematic setup, we compute the amplitude for 
$\mathrm{Ps}^- \to e^- Z^*$ 
along the $z$ axis using {\tt FeynArts}
~\cite{Hahn2001}
and {\tt FeynCalc}~\cite{Shtabovenko2016}. The resulting amplitudes are listed below:
\begin{eqnarray}
\large
	\mathcal{M}_1&=&\frac{i e^3\varepsilon^{*\nu}(q)\left(\bar{v}(p_2)\gamma_{\mu}u(p_1) \right)\left(\bar{u}(k_3)\gamma^\mu (\slashed{p}_3-\slashed{q}+m)\gamma_{\nu}(v_\ell-a_\ell\gamma_{5})u(p_3) \right)  }{(q+k_3-p_3)^2((q-p_3)^2-m^2)}\;,\notag\\
	\mathcal{M}_2&=&\frac{i e^3\varepsilon^{*\nu}(q)\left(\bar{v}(p_2)\gamma_{\mu}u(p_1) \right)\left( \bar{u}(k_3)\gamma^{\mu}(v_\ell-a_\ell \gamma_{5})(\slashed{q}+\slashed{k}_3+m)\gamma_{\nu}u(p_3)\right)  }{(q+k_3-p_3)^2((-q-k_3)^2-m^2)}\;,\notag\\
	\mathcal{M}_3&=&-\frac{i e^3\varepsilon^{*\nu}(q)\left(\bar{v}(p_2)\gamma_\mu u(p_3) \right)\left(\bar{u}(k_3)\gamma^\mu (\slashed{k}_3-\slashed{p}_2-\slashed{p}_3+m)\gamma_{\nu}(v_\ell-a_\ell\gamma_5)u(p_1) \right)  }{(p_2+p_3)^2((-k_3+p_2+p_3)^2-m^2)}\;,\notag\\
	\mathcal{M}_4&=&-\frac{i e^3 \varepsilon^{*\nu}(q)\left( \bar{v}(p_2)\gamma_{\mu}u(p_3)\right)\left(\bar{u}(k_3)\gamma_{\nu}(v_\ell-a_\ell\gamma_{5})(\slashed{q}+\slashed{k}_3+m)\gamma^\mu u(p_1) \right)  }{(-p_2-p_3)^2((-q-k_3)^2-m^2)}\;,\notag\\
	\mathcal{M}_5&=&\frac{i e^3\varepsilon^{*\nu}(q)(\bar{u}(k_3)\gamma_\mu u(p_3))(\bar{v}(p_2)\gamma^\mu(\slashed{k}_3-\slashed{p}_2-\slashed{p}_3+m)\gamma_\nu(v_\ell-a_\ell\gamma_{5})u(p_1))}{(k_3-p_3)^2((-k_3+p_2+p_3)^2-m^2)}\;,\notag
	\\
	\mathcal{M}_6&=&-\frac{i e^3\varepsilon^{*\nu}(q)(\bar{u}(k_3)\gamma_\mu u(p_1))(\bar{v}(p_2)\gamma_\nu(v_\ell-a_\ell\gamma_{5})(\slashed{q}-\slashed{p}_2+m)\gamma^\mu u(p_3))}{(q-p_2-p_3)^2((p_2-q)^2-m^2)}\;,\notag\\
	\mathcal{M}_7&=&\frac{i e^3 \varepsilon^{*\nu}(q)(\bar{u}(k_3)\gamma_\mu u(p_3))(\bar{v}(p_2)\gamma_\nu(v_\ell-a_\ell\gamma_{5})(\slashed{q}-\slashed{p}_2+m)\gamma^\mu u(p_1))}{(k_3-p_3)^2((p_2-q)^2-m^2)}\;,\notag\\
	\mathcal{M}_8&=&-\frac{i e^3\varepsilon^{*\nu}(q)(\bar{u}(k_3)\gamma_\mu u(p_1))(\bar{v}(p_2)\gamma^\mu (\slashed{p}_3-\slashed{q}+m)\gamma_\nu (v_\ell-a_\ell\gamma_5)u(p_3))}{(q-p_2-p_3)^2((q-p_3)^2-m^2)}\;.\label{amplitude}
\end{eqnarray}
Here 
%$\text{G}_\text{F} \simeq 1.166 \times 10^{-5}\,\mathrm{GeV}^{-2}$ denotes the Fermi constant~\cite{Marciano1999}, $\alpha$ is the fine-structure constant, and 
$\varepsilon^{*\nu}(q)$ represents the polarization vector of the massive $Z$ boson with four-momentum $q$. 
The effective vector and axial-vector couplings induced by $Z$ boson exchange for $\ell=\mu$ are given by $
v_\ell = \frac{1}{4} - \sin^2\theta_W\;,
a_\ell = \frac{1}{4}$, 
where $\theta_W$ is the weak mixing angle. Numerically, 
$\sin^2\theta_W \simeq 0.238$~\cite{czarnecki2005}.
\\
Based on the structure of the $e^-e^+$ vertex in Eq.~\ref{amplitude}, the amplitudes naturally separate into ortho- and para-positronium-like contributions. Specifically, $\mathcal{M}_1$--$\mathcal{M}_4$ [Figs.~\ref{fig:dia}(a)--(d)] arise from spin-triplet ($S=1$) annihilation, whereas $\mathcal{M}_5$--$\mathcal{M}_8$ [Figs.~\ref{fig:dia}(e)--(h)] arise from spin-singlet ($S=0$) annihilation. The amplitudes $\mathcal{M}_3$, $\mathcal{M}_4$, $\mathcal{M}_7$, and $\mathcal{M}_8$ are obtained from $\mathcal{M}_1$, $\mathcal{M}_2$, $\mathcal{M}_5$, and $\mathcal{M}_6$, respectively, by exchanging the two identical electrons.

\subsection{Specific Spin Configuration}\label{spin-config}

The complete wave function of Ps$^-$ can be determined by the variational principle~\cite{bhatia1983}. The spatial part of Ps$^-$ in the ground state is symmetric with respect to the e$^-$ pair. For the total wave function to be anti-symmetric, the two electrons  must be in a spin singlet state. The positron can be in spin-up or spin-down, and hence the total spin-wave function of $\mathrm{Ps}^-$ is $\frac{1}{\sqrt{2}}\left(\uparrow\downarrow -\downarrow\uparrow\right)\left(\uparrow,\downarrow\right)$. This reduces the number of spin configurations allowed for the decay.

In the ground state, if we take the positron to be in either a spin-up or a spin-down state, its annihilation with one of the electrons in the $\uparrow\downarrow$ configuration can proceed through either a spin-triplet ($S=1$, analogous to ortho-positronium) or a spin-singlet ($S=0$, analogous to para-positronium) state. 
In general, the spin of this annihilating $e^+e^-$ pair then combines with the spin of the remaining electron ($S=\tfrac{1}{2}$). The addition of angular momenta yields $
1 \otimes \tfrac{1}{2} = \tfrac{1}{2}\;, \tfrac{3}{2}$, $
0 \otimes \tfrac{1}{2} = \tfrac{1}{2}$.   Therefore, the total spin of the system can be either $J=\tfrac{1}{2}$ or $J=\tfrac{3}{2}$.
Using the standard Clebsch–Gordan coefficients, the initial state - labeled by total angular momentum $J$ and its magnetic projection $m_j$ - can be decomposed in terms of the final particle spin states~\cite{ParticleDataGroup:2024cfk}. 

\begin{table}[htp]
\renewcommand{\arraystretch}{1.9}
\large
	\begin{center}
		\begin{tabular}{|c|c|c|c|c|c|c|}
			\hline
			$m_j$&
			\multicolumn{3}{|c|}{Incoming} & 
			\multicolumn{2}{|c|}{Outgoing}& Amplitude \\ \hline
			% 4 rows combined
			\multirow{4}{*}{$\frac{1}{2}$}
			 &
			$e^-$ & $e^+$ & $e^-$ & $Z^*$ & $e^-$&\\
			\cline{2-7}
			
			\multicolumn{1}{|c|}{}&
			\ua & \ua & \da & $+1$ & \da& \large$\frac{e^3\sqrt{E_e+m}(a_\ell\sqrt{E_e^2-m^2}(5E_e-9m)+v_\ell(5E_e-11m)(E_e-m))}{4m^{3/2}(E_e-m)^2}$ \\
			\cline{2-7}
			
			\multicolumn{1}{|c|}{}
            &
			\da & \ua & \ua & $+1$ & \da& \large$-\frac{e^3\sqrt{E_e+m}(a_\ell\sqrt{E_e^2-m^2}(5E_e-9m)+v_\ell(5E_e-11m)(E_e-m))}{4m^{3/2}(E_e-m)^2}$ \\
			\cline{2-7}
			
			\multicolumn{1}{|c|}{}  &
			\ua & \ua & \da & 0 & \ua& \large$\frac{e^3\sqrt{E_e+m}(-a_\ell\sqrt{E_e^2-m^2}(E_e+5m)+v_\ell(E_e^2-3mE_e+4m^2)}{2\sqrt{2q^2}m^{3/2}(E_e-m)}$ \\
			\cline{2-7}
			
			\multicolumn{1}{|c|}{}  &
			\da & \ua & \ua & 0 & \ua&\large$-\frac{e^3\sqrt{E_e+m}(-a_\ell\sqrt{E_e^2-m^2}(E_e+5m)+v_\ell(E_e^2-3mE_e+4m^2)}{2\sqrt{2q^2}m^{3/2}(E_e-m)}$  \\
			\cline{2-7}

			\hline
			
			% again, 3 rows combined
			\multirow{4}{*}{$-\frac{1}{2}$}
		&\da&\da&\ua&-1&\ua& -\large$\frac{e^3\sqrt{E_e+m}(v_\ell(E_e-m)^2-a_\ell(E_e-3m)\sqrt{E_e^2-m^2})}{2(E_e-m)^2m^{3/2}}$ \\
			\cline{2-7}
			
			\multicolumn{1}{|c|}{} &\ua&\da&\da&-1&\ua&\large$\frac{e^3\sqrt{E_e+m}(v_\ell(E_e-m)^2-a_\ell(E_e-3m)\sqrt{E_e^2-m^2})}{2(E_e-m)^2m^{3/2}}$  \\ 
			
			\cline{2-7}
			\multicolumn{1}{|c|}{} &\da&\da&\ua&0&\da&\large$\frac{e^3\sqrt{E_e+m}(a_\ell\sqrt{E_e^2-m^2}(E_e+m)+v_\ell(E_e^2-3mE_e+4m^2))}{\sqrt{2q^2}m^{3/2}(E_e-m)}$  \\

			\cline{2-7}
			\multicolumn{1}{|c|}{} &\ua&\da&\da&0&\da&\large$-\frac{e^3\sqrt{E_e+m}(a_\ell\sqrt{E_e^2-m^2}(E_e+m)+v_\ell(E_e^2-3mE_e+4m^2))}{\sqrt{2q^2}m^{3/2}(E_e-m)}$   \\ 
			\hline
		\end{tabular}
	\end{center}
	\caption{The amplitudes for the various spin configurations of the annihilating $e^+e^-$ pair, together with the spin states of the remaining initial- and final-state $e^-$ and the polarization of the $Z$ boson, are shown. The $0\otimes \frac{1}{2}$ and $1\otimes \frac{1}{2}$ couplings give rise to states with total angular momentum $j=\frac{1}{2} ,\frac{3}{2}$ and projections $m_j=\pm\frac{1}{2},,\pm\frac{3}{2}$. However, the amplitudes for $m_j=\pm\frac{3}{2}$ vanish, so that only the spin-singlet electron-pair configuration contributes.
}\label{spin-configurations}
\end{table}

Based on the aforementioned considerations, together with the spinor combinations defined in Appendix~\ref{spinor}, we compute the scattering amplitudes for all possible spin configurations. In particular, we consider the electron pair in both spin-singlet and spin-triplet states, as well as all spin states of the final-state electron and the longitudinal and transverse polarizations of the $Z$ boson. The resulting amplitudes for the various spin and $Z$ boson polarization assignments are summarized in Table~\ref{spin-configurations}. As evident from the Table~\ref{spin-configurations}, only the spin-singlet electron configuration gives a nonvanishing contribution, whereas the amplitudes for all other spin configurations vanish.

Collecting the results for the spin-$\uparrow$ and spin-$\downarrow$ final-state electron configurations, together with the longitudinal and transverse polarizations of the $Z$ boson, the decay amplitudes for $\mathrm{Ps}^- \to e^- Z^*$ are given by
  \begin{align}
  	\mathcal{M}^{free}_{\mathrm{Ps}^-\rightarrow e^- Z^*}
  	&=\begin{cases}
    \huge
    \frac{3\sqrt{2}e^3\sqrt{E_e+m}(a_\ell\sqrt{E_e^2-m^2}(5E_e-9m)+v_\ell(5E_e-11m)(E_e-m))}{4m^{3/2}(E_e-m)^2},&\varepsilon^*(q)=+1, e^- =\; \downarrow
  		\\
  \huge		\frac{3\sqrt{2}e^3\sqrt{E_e+m}(v_\ell(E_e-m)^2-a_\ell(E_e-3m)\sqrt{E_e^2-m^2})}{2m^{3/2}(E_e-m)^2},&\varepsilon^*(q)=-1, e^-=\;\uparrow
  		\\
 \huge
 \frac{3e^3\sqrt{E_e+m}(-a_\ell\sqrt{E_e^2-m^2}(E_e+5m)+v_\ell(E_e^2-3mE_e+4m^2))}{2m^{3/2}\sqrt{q^2}(E_e-m)},&\varepsilon^*(q)=0, e^-=\;\uparrow
  		\\
 \huge 
 \frac{3e^3\sqrt{E_e+m}(a_\ell\sqrt{E_e^2-m^2}(E_e+m)+v_\ell(E_e^2-3mE_e+4m^2))}{\sqrt{q^2}m^{3/2}(E_e-m)},&\varepsilon^*(q)=0, e^-\;=\downarrow\;.
  	\end{cases}
    \label{ampl-cases}
  \end{align}
%where the factor of 3 is due to the fact that all amplitudes contribute three times. 
Eq.~\ref{ampl-cases} represents the projections of the electroweak current onto the allowed polarizations of the $Z^*$ boson. It can be seen that amplitudes naturally separate into vector $\left(v_\ell\right)$ and axial vector $\left(a_\ell\right)$ contributions.
For the transverse polarization, the electron spin configuration is always opposite to the helicity of the $Z^*$ boson polarization, reflecting the chiral structure of the electroweak interaction. On the other hand, both electrons' spin states contribute in longitudinal polarization.

These free state amplitudes are related to the bound state by~\cite{Peskin}:
\begin{align}
\mathcal{M}^{Bound}=\sqrt{2M}\frac{1}{\sqrt{2E_{e^-}}\sqrt{2E_{e^+}}\sqrt{2E_{e^-}}}\Psi\left(0,0,0\right)\mathcal{M}^{free}\;,\label{free-bound}
\end{align}
where $M$ is the mass of the positronium ion, $E_{e^-}=E_{e^+}=m$, and the value of  $\Psi(0,0,0)$ is given in Eq. (\ref{coel-prob}) \cite{Aslam:2021uqu}.

We use the standard three-body decay rate formula
\begin{align}	\label{eq:3bodydecay}
	\Gamma_{\mathrm{Ps}^-\to e^-\nu_\mu\bar{\nu}_\mu}=\frac{1}{2M}\int d\phi_3(p_1+p_2+p_3;k_1+k_2+k_3)\frac{3|\Psi(0,0,0)|^2}{4m^2}\frac{1}{g}\sum_{\text{pol.}/\text{spin}}|\mathcal{M}^{free}|^2\;.
\end{align}
By decomposing the three-body phase space into a product of two two-body phase spaces, corresponding to $\mathrm{Ps}^- \to e^- Z^*$ followed by $Z^* \to \nu_\mu \bar{\nu}_\mu$, the decay rate can be expressed as (see Appendix~\ref{app:3bodydecay} for details):
\begin{equation}
\Gamma_{\mathrm{Ps}^-\to e^-\nu_\mu\bar{\nu}_\mu}=\frac{4G_F^2}{2\pi^2\alpha}\int\frac{dq^2}{2\pi}q^2\left(\frac{1}{2M}\int d\phi_2(3p,q,k_3)\frac{3|\Psi(0,0,0)|^2}{4m^2}\frac{1}{3g}\sum_{\text{pol.}/\text{spin}}|\mathcal{M}^{free}_{\mathrm{Ps}^-\to e^-Z^*}|^2 \right)\;.\label{drate-1}
\end{equation}

In Eq.~(\ref{ampl-cases}), we present the decay amplitudes for the various spin and polarization configurations, which are mutually orthogonal. Consequently, using the 2-body decay kinematics given in Eq. (\ref{eq:2kin}), the total decay rate can be obtained by evaluating the contribution from each spin configuration separately and summing over their squares, using Eq.~(\ref{drate-1}). 

%We use the standard two-body decay rate formula, two-body kinematics in Appendix~\ref{eq:2kin}, and Eq.~\ref{eq:3bodydecay} to compute three body decay rate.
%Since we consider a small binding energy of Ps$^-$, we consider the mass of Ps as 3m-$\epsilon$, where $\epsilon$ is very small compared to the rest mass of e$^-$. We use the limit of integration $0\leq q^2<4m^2$. This is because at $q^2=4m^2$, the internal propagator becomes on shell, and thus break the condition for perturbation theory. 
We have decomposed the phase space into a product of two-body phase spaces corresponding to $\mathrm{Ps}^- \to e^- Z^*$ followed by $Z^* \to \nu_\mu \bar{\nu}_\mu$, and integrated over the invariant mass squared of the neutrino pair, $q^2 = (k_1 + k_2)^2$. The kinematically allowed range of $q^2$ is determined entirely by energy–momentum conservation. In the present case, taking the neutrinos to be massless and approximating the positronium mass as $M = 3m - \epsilon$, with binding energy $\epsilon \ll m$, we obtain
\begin{equation}
0 \le q^2 \le (M - m)^2 = (2m - \epsilon)^2 \simeq 4m^2 - 4m\epsilon\;.
\end{equation}
Thus, the upper limit follows purely from kinematics and lies slightly below $4m^2$ due to the small binding energy of $\mathrm{Ps}^-$.
Finally, integrating over $q^2$, the total decay rate for various spin configurations comes out to be:
\begin{align}
\label{eq:lwaresult}
\Gamma_{\mathrm{Ps}^- \to e^- \nu_\mu\bar{\nu}_\mu}\approx1\times10^{-20}\;\text{s}^{-1}\;.
\end{align} 
Compared with the electromagnetic decay $\mathrm{Ps}^- \to e^- \gamma$, the branching ratio for the weak decay $\mathrm{Ps}^- \to e^- \nu_\mu \bar{\nu}_\mu$ is highly suppressed, as expected for a weak-interaction process. We obtain
\begin{align}
\mathrm{Br}(\mathrm{Ps}^- \to e^- \nu_\mu \bar{\nu}_\mu)
= \frac{\Gamma(\mathrm{Ps}^- \to e^- \nu_\mu \bar{\nu}_\mu)}
       {\Gamma(\mathrm{Ps}^- \to e^- \gamma)}
\approx 10^{-19}\;.
\end{align}
The extreme suppression of the branching ratio can be understood parametrically: the weak decay amplitude is proportional to the Fermi constant $G_F$, while the dominant electromagnetic decay proceeds via a single photon. Therefore, the branching ratio scales roughly as
\[
\mathrm{Br}(\mathrm{Ps}^- \to e^- \nu_\mu \bar{\nu}_\mu) \sim \frac{G_F^2 m^4}{\alpha} \;.
\]
This parametric estimate highlights why the weak channel is so tiny compared with the electromagnetic decay.
This is comparable to the branching ratio of the weak-decay of ortho-positronium, $\text{o-}\mathrm{Ps}\to \gamma\nu\bar\nu$ estimated in~\cite{Pokraka:2016jgy, Bernreuther}:
\begin{align}
\mathrm{Br}(\mathrm{Ps} \to \gamma \nu \bar{\nu})
=\frac{\Gamma(\text{o-}\mathrm{Ps}\to \gamma \nu \bar{\nu})}{\Gamma(\text{o-}\mathrm{Ps}\to3\gamma)}\approx 10^{-19}\;.
\end{align}

The similarity between the weak-to-electromagnetic decay ratios in $\mathrm{Ps}$ and the $\mathrm{Ps}^-$ follows from the factorization properties of nonrelativistic quantum electrodynamics (NRQED). In annihilation processes, the decay widths factorize into a short-distance coefficient and the $e^{+}e^{-}$ contact probability $\langle \delta^{3}(\mathbf{r}_{e^{+}e^{-}}) \rangle$. Since this bound-state factor is common to both channels, it cancels in the ratio, leaving the result determined primarily by the relative strength of the weak and electromagnetic interactions.

\section{General spin Configuration}
The second method follows the standard approach in perturbative quantum field theory. One first computes the decay amplitude, multiplies it by its complex conjugate, and then averages over the spins of the initial-state particles while summing over the spins of the final-state particles. The spin sums can be expressed in terms of traces over Dirac matrices using the completeness relations for spinors, a procedure commonly referred to as the Casimir trick. The squared amplitude is therefore given by
\begin{equation}
|\mathcal{M}|^{2}
=
\frac{1}{8}
\sum_{\text{all spins}}
|M|^{2}\;,
\label{eq:1}
\end{equation}
where the factor of $\frac{1}{8}$ ensures spin averaging over the initial state particles.

Using \texttt{FeynArts}~\cite{Hahn2001}, we generated all contributing Feynman diagrams and constructed the corresponding amplitudes, which are for the two-body decay process $\textrm{Ps}^-\to e^-Z^*$ in Eq. (\ref{amplitude}). To get the corresponding amplitude for the three-body decay process $\textrm{Ps}^-\to e^-\nu_\mu\bar{\nu}_\mu$, along with appropriate constants, we simply replace $\varepsilon^{*\nu}$ with the corresponding weak-neutral current: 
\begin{equation}
  J^\nu(k_1,k_2)=\bar{u}(k_1)\gamma^\nu(1-\gamma_5)v(k_2)\;, 
\end{equation}
Since there are 8 diagrams for this process, the evaluation of $|M|^{2}$ involves 64 terms in total. The fermion spin sums and Dirac traces are computed with the Mathematica package \texttt{FeynCalc}~\cite{Shtabovenko2016}. The resulting squared amplitude is:
\begin{eqnarray}
|\mathcal{M}^{free}|^2&=&\frac{e^4G_F^2\;q^2(16m^2-q^2)}{70m^4(4m^2-q^2)^3(8m^2+q^2)^6}P(m^2,q^2)
\end{eqnarray}
Here,
\begin{eqnarray}
  \nonumber
  P(m^2,q^2)=\sum_{n=0}^{9}c_nm^{18-2n}q^{2n},
\end{eqnarray}
\\
\begin{table}[htp]
\centering
\renewcommand{\arraystretch}{1.3}
\setlength{\tabcolsep}{8pt}

\begin{tabular}{|c|cccccccccc|}
\hline
$n$   & 0 & 1 & 2 & 3 & 4 & 5 & 6 & 7 & 8 & 9 \\
\hline
$c_n$ & $3\times10^{11}$ 
      & $-10^{11}$ 
      & $7\times10^{10}$ 
      & $10^{10}$ 
      & $2\times10^{9}$ 
      & $5\times10^{7}$ 
      & $-5\times10^{6}$ 
      & $-2\times10^{6}$ 
      & $-84730$ 
      & $-5225$ \\
\hline
\end{tabular}

\caption{Coefficients $c_n$ appearing in the polynomial
\(
P(x)=\sum_{n=0}^{9} c_n x^n
\).
}
\label{tab:coefficients}
\end{table}
In term of a dimensionless number $x=\frac{q^2}{m^2}$, the amplitude square is:
\begin{eqnarray}
    |\mathcal{M}^{free}|^2=\frac{ax(16-x)}{(4-x)^3(8+x)^6}\sum_{n=0}^{9}c_nx^n,
\end{eqnarray}
where $a=\frac{e^4G_F^2}{70}$, and coefficients $c_2$ are written in Table \ref{tab:coefficients}.
Relating the free-state amplitude to the bound-state amplitude via Eq.~(\ref{free-bound}), and employing the three-body kinematics summarized in Appendix~\ref{eq:3kin}, the phase-space decomposition in Appendix~\ref{app:phase}, and the standard three-body decay formula, we obtain
\begin{align}
\Gamma_{\mathrm{Ps}^-\to e^- \nu_\mu \bar{\nu}_\mu}
=
0.15\, G_F^{\,2}\,\alpha^{8}\, m^{5}
\approx 1 \times 10^{-20}\ \mathrm{s}^{-1}.
\end{align}
This result is consistent with Eq.~(\ref{eq:lwaresult}) and confirms that only the spin-singlet electron configuration in $\mathrm{Ps}^-$ contributes a non-vanishing rate to the weak decay under consideration.

\section{Conclusion}\label{conclusions}

The electromagnetic decay $\mathrm{Ps}^- \to e^- \gamma$ has been studied in the literature. 
In this work, we consider the electroweak annihilation process $e^+e^- \to \gamma Z^*$ 
in the presence of a third electron that absorbs the photon, while the virtual $Z$ boson 
decays into a neutrino-antineutrino pair. At the bound-state level, this corresponds to 
the weak decay $\mathrm{Ps}^- \to e^- \nu_\mu \bar{\nu}_\mu$.

We compute the decay rate of $\mathrm{Ps}^-$ using two independent methods. In the first 
approach, we consider all spin configurations of the constituents and apply the standard 
electroweak rules to express the amplitudes in terms of Dirac spinors, which are reduced 
to traces of $\gamma$-matrices. By decomposing the three-body phase space as 
$\mathrm{Ps}^- \to e^- Z^*\!\left(\to \nu_\mu \bar{\nu}_\mu\right)$, we find that a 
nonvanishing rate arises only when the electron pair in $\mathrm{Ps}^-$ is in a spin-singlet state.

In the second approach, we generate the leading-order diagrams using 
\texttt{FeynArts} \cite{Hahn2001} and evaluate the squared amplitude with 
\texttt{FeynCalc} \cite{Shtabovenko2016}. Both methods yield the same decay rate. 
The ratio of the electroweak to electromagnetic decay rates scales as $\tfrac{G_F^2 m^4}{\alpha}$, yielding a branching ratio comparable to that of $\mathrm{o}$-$\mathrm{Ps}\to\gamma\nu\bar{\nu}$. This follows from the identical photon-exchange topology in the weak and electromagnetic decays of $\mathrm{Ps}^-$ and the cancellation of the common bound-state factor in the decay-rate ratio.

\section*{Acknowledgments}
M. J. A would like to thank Prof. A. Czarnecki for introducing him to this subject for the first time in 2019. During the preparation of this study, the authors used ChatGPT 5.2 (free version) for the purposes of language editing and to correct grammatical mistakes. The authors have reviewed and edited the output and take full responsibility for the content of this work.
\appendix
\numberwithin{equation}{section}
\section{Phase space}\label{app:phase}
	The $N$ body phase space can be written as:
	\begin{align}
		d\phi_N(p^\mu;k_1,....k_N)=(2\pi)^4\delta^4(p^\mu-\sum_{i=1}^Nk_i^\mu)\prod_{i=1}^{N}\frac{d^3k_i}{(2\pi)^3}\frac{1}{2E_i}\;,
	\end{align}
	where $p^\mu$ is the four momentum of the initial particles. $k_i^\mu$ and $E_i$ are the momentum and energy of the $i^{th}$ particle in final state and the factor $\frac{d^3k_i}{(2\pi)^3}\frac{1}{2E_i}$ is the Lorentz invariant that is used for normalization. In this section, we discuss how we can decompose a three-body phase space into two two-body phase spaces.
	
\subsection{Decomposition of three-body phase space}
	
	The three body phase space can be written as,
	\begin{align}
		\label{eq:3phase}
		\int d\phi_3(P;k_1,k_2,k_3)=\int \prod_{i=1}^{3}\frac{d^3k_i}{(2\pi)^3 2E_i}(2\pi)^4\delta^4(P-k_1-k_2-k_3)\;.
	\end{align}
Where $P=p_1+p_2+p_3$ is the total initial momentum. We can write the unity in terms of an integral as
	\begin{align}
		\label{eq:iden}
	&1=\int\frac{d^3q}{(2\pi)^32E_q}\frac{dq^2}{2\pi}(2\pi)^4\delta^4(q-k_1-k_2)\;.
\end{align}
Here $q=k_1+k_2$ is the four momentum of the virtual $Z$ boson.
By substituting Eq~\ref{eq:iden} in Eq~\ref{eq:3phase} , we get
	\begin{align}
	\nonumber
	\int d\phi_3(P;k_1,k_2,k_3)=&\int \frac{d^3k_1}{(2\pi)^32E_1}\frac{d^3k_2}{(2\pi)^32E_2}\frac{d^3k_3}{(2\pi)^32E_3}(2\pi)^4\delta^4(P-k_1-k_2-k_3)\\
	\nonumber
	\times &\int \frac{d^3q}{(2\pi)^32E_q}\frac{dq^2}{2\pi}(2\pi)^4\delta^4(q-k_1-k_2)\\
	=&\int \frac{dq^2}{2\pi}d\phi_2(P;q,k_3)d\phi_2(q;k_1,k_2)
\end{align}
In terms of the K\"all\'en function $\lambda\left(a,b,c\right)= a^2 + b^2 + c^2 - 2ab - 2ac - 2bc.$, the phase space becomes 
\begin{align}
    \int d\phi_3 (P;k_1,k_2,k_3)=\int \frac{dq^2}{2\pi}\int \frac{d\theta_3}{2}\frac{d\phi_3}{2\pi}\frac{\widetilde{\beta}\left(P^2,q^2,k_3^2\right)}{8\pi}\int \frac{d\theta_{12}}{2}\frac{d\phi_{12}}{2\pi}\frac{\widetilde{\beta}\left(q^2,k_1^2,k_2^2\right)}{8\pi}\;,
\end{align}
where
\begin{align}
    \Tilde{\beta}(s,s_1,s_2)=\sqrt{\frac{\lambda}{s^2}}=\sqrt{1-\frac{2(s_1+s_2)}{s}+\frac{(s_1-s_2)^2}{s^2}}
\end{align}
%\textcolor{red}{Nishat: Is it correct now?}%
\subsection{Factorization of three body decay rate for $\mathrm{Ps}^-\to e^-\nu_\mu\bar{\nu}_\mu$}\label{app:3bodydecay}
	
	The Feynman graphs relevant to the decay $\mathrm{Ps}^-\rightarrow e\nu_\mu\bar{\nu}_\mu$ are drawn in Figure \ref{fig:f1}. For any of these diagrams, the LO amplitude is of the form
	\begin{align}
		\mathcal{M}=\frac{2G_FJ^{\nu*} X_\nu}{\sqrt{2\pi\alpha}}\;,\label{amp-appendix}
	\end{align}
    where
    \begin{equation}
    X_\nu=\left( \bar{v}(p_2)ie\gamma_\mu u(p_1)\right)\frac{1}{q_\gamma^2}\left( \bar{u}(k_3)ie\gamma^\mu\frac{i(\slashed{q_e}+m)}{q_e^2-m^2}ie\gamma_\nu(v_\ell-a_\ell\gamma_5)\right)\;,
    \end{equation}
    and
\begin{equation}
 J^{\nu *}(k_1,k_2)=\left( \bar{u}(k_1)\gamma^\nu(1-\gamma_5)v(k_2)\right)\;,
 \end{equation}
is the neutrino weak neutral current.

We use the standard three-body decay rate formula:
\begin{align}
	\label{eq:3bodydecay}
	\Gamma_{\text{Ps}^-\rightarrow e\nu\bar{\nu}}=\frac{1}{2M}\int d\phi_3(p_1+p_2+p_3;k_1+k_2+k_3)\frac{3|\Psi(0,0,0)|^2}{4m^2}\frac{1}{g}\sum_{\text{pol.}/\text{spin}}|\mathcal{M}|^2\;,
\end{align}
where $M$ is the mass of $\mathrm{Ps}^-$ and $g$ is the number of $\mathrm{Ps}^-$ polarizations of in initial states. Decomposing the three body phase space given in Appendix~(\ref{app:phase}) [c.f. Eq. (\ref{eq:3phase})] and substituting into Eq~\ref{eq:3bodydecay}, we get
\begin{align}
	\Gamma_{\mathrm{Ps}^-\to e^-\nu_\mu\bar{\nu}_\mu}=\frac{1}{2m_{\text{Ps}^-}}\int \frac{dq^2}{2\pi}(3p;q,k_3)\frac{3|\Psi(0,0,0)|^2}{4m^2g}\frac{4G_F^2}{2\pi\alpha}X_\nu X_\alpha^*\int d\phi_2(q;k_1,k_2)J^{\nu *} J^{\alpha}\;.
\end{align}
	The neutrinos contribution to the decay rate gives the sum over the polarization of the massive $Z$ boson, \textit{i.e.,} 
    \begin{align}
        \int d\phi_2(q;k_1,k_2)J^{\nu*} J^{\alpha}=\frac{q^2}{3\pi}\sum_{\text{pol}./\text{spin}}\varepsilon^{\nu*}(q)\varepsilon^{\alpha}(q)\;.
    \end{align}

    We obtain the $\mathrm{Ps}^-\to e^-\nu_\mu\bar{\nu}_\mu$ rate in terms of $\mathrm{Ps}^-\to e^-Z^*$.
	\begin{align}
		\nonumber
		\Gamma_{\mathrm{Ps}^-\to e^-\nu_\mu\bar{\nu}_\mu}=&\frac{1}{2M}\int \frac{dq^2}{2\pi}d\phi_2(2p;q,k_3) \frac{3|\Psi(0,0,0)|^2}{4m^2g}\frac{4G_F^2}{2\pi\alpha}X_\nu X_\alpha^*\frac{q^2}{3\pi}\sum_{\text{pol.}/\text{spin}}\varepsilon^{\nu*}(q)\varepsilon^{\alpha}(q)\\
		\nonumber
		=&\frac{4G_F^2}{2\pi^2\alpha}\int\frac{dq^2}{2\pi}q^2\left(\frac{1}{2M}\int d\phi_2(2p,q,k_3)\frac{3|\Psi(0,0,0)|^2}{4m^2}\frac{1}{3g}\sum_{\text{pol.}/\text{spin}}|\mathcal{M}_{\text{Ps}^-\rightarrow eZ^*}|^2 \right) \\
		=&\frac{4G_F^2}{2\pi^2\alpha}\int \frac{dq^2}{2\pi}q^2\Gamma_{\mathrm{Ps}^-\to e^-Z^*}\;.
	\end{align}
    
\section{Kinematics}\label{app:kinematics}
	For the two-body $\mathrm{Ps}^-\to e^- Z^*$ decay, we assume that the electron moves along the +$z$-axis. To conserve the three momentum, the $Z$ boson should move along the $-z$-axis. Using conservation of three momentum and energy, one can obtain:
		\begin{equation}
    \label{eq:2kin}
			E_e=\frac{10m^2-q^2}{6m},\quad       E_q=\frac{8m^2+q^2}{6m}\;,\quad
	|\vec{k}_3|=\sqrt{E_e^2-m^2}\;.
\end{equation}

Similarly, for $\mathrm{Ps}^-\to e^-\nu_\mu\bar{\nu}_\mu$ three body decay rate, the kinematical relations are:
\begin{equation}
\label{eq:3kin}
	E_{e}=\frac{10m^{2}-q^{2}}{6m},\quad E_{\nu}=|\vec{p}_\nu|=\frac{(\frac{q^2+8m^2}{6m})^2-|\vec{k}_3|^{2}}{2(\frac{q^2+8m^2}{6m})}\left(1-\frac{|\vec{k}_3|\cos\theta}{\frac{q^2+8m^2}{6m}}+\frac{(|\vec{k}_3|\cos\theta)^2}{(\frac{q^2+8m^2}{6m})^2} \right)\;, 
\end{equation}
\begin{equation}
	|\vec{k}_3|=\sqrt{E_e^2-m^2},\quad
	E_{\bar{\nu}}=|\vec{p}_{\bar{\nu}}|=3m-E_e-E_\nu\;.
\end{equation}
	\subsection{Combination of spinors}
    \label{spinor}
	Using the definition of Dirac spinors \cite{Peskin}, we can write them for the spin-up and -down outgoing electron. In terms of the $\gamma$ matrices, we used the following form in our calculation \cite{Aslam:2021uqu, Aslam2023}:
	\begin{align}	u_{\uparrow}\left(p\right)\bar{u}_{\uparrow}\left(k_3\right)  =&\frac{\sqrt{2m(E_e+m)}}{4}\left(1+\gamma^{0}\right)\left(\gamma^{5}+\gamma^{3}\right)\left(\gamma^{5}-\frac{|\vec{k}_3|}{E_e+m}\right)\;,\label{eq:psi4}
	\\
	u_{\downarrow}\left(p\right)\bar{u}_{\uparrow}\left(k_3\right)  =&-\frac{\sqrt{2m(E_e+m)}}{4}\left(1+\gamma^{0}\right)\left(\frac{|\vec{k}_3|}{E_e+m}+\gamma^{5}\right)\left(\gamma^{1}-i\gamma^{2}\right)\;,\label{eq:psi5}
\\
	u_{\uparrow}\left(p\right)\bar{u}_{\downarrow}\left(k_3\right)   =&\frac{\sqrt{2m(E_e+m)}}{4}(1+\gamma^{0})(\gamma^{1}+i\gamma^{2})\left(\gamma^{5}+\frac{|\vec{k}_3|}{E_e+m}\right)\;,\label{eq:pi6}
\\
	u_{\downarrow}\left(p\right)\bar{u}_{\downarrow}\left(k_3\right)  =&\frac{\sqrt{2m(E_e+m)}}{4}(\gamma^{5}-\gamma^{3})(1-\gamma^{0})\left(\gamma^{5}+\frac{|\vec{k}_3|}{E_e+m}\right)\;.\label{eq:pi7}
\end{align}

For the initial states, the spinor combinations are:
\begin{align*}
	u_{\uparrow}(p)\bar{v}_{\uparrow}(p) & =2m\frac{1+\gamma^{0}}{2}\frac{\gamma^{1}+i\gamma^{2}}{2},\quad u_{\uparrow}(p)\bar{v}_{\downarrow}(p)=-2m\frac{1+\gamma^{0}}{2}\frac{\gamma^{5}+\gamma^{3}}{2},\\
	u_{\downarrow}(p)\bar{v}_{\downarrow}(p) & =-2m\frac{1+\gamma^{0}}{2}\frac{\gamma^{1}-i\gamma^{2}}{2},\quad u_{\downarrow}(p)\bar{v}_{\uparrow}(p)=2m\frac{1+\gamma^{0}}{2}\frac{\gamma^{5}-\gamma^{3}}{2}.
\end{align*}

\end{document}